\documentclass[aps,onecolumn,secnumarabic,nobalancelastpage,amsmath,amssymb,nofootinbib]{revtex4}

\usepackage{graphics}      
\usepackage{graphicx}      
\usepackage{subfig}
\usepackage{url}           
\usepackage{bm}            
\usepackage[utf8]{inputenc}
\usepackage[english]{babel}
\usepackage{booktabs}
\usepackage{comment}
\usepackage{adjustbox}

\begin{document}
\title			{Equatorial circular motion in a Kerr-like spacetime}
\author{Diego Solano Alfaro$^1$}
\author{Francisco Frutos Alfaro$^{1, 2}$}
\email          {diego.solanoalfaro@ucr.ac.cr, francisco.frutos@ucr.ac.cr}
\homepage       {http://www.fisica.ucr.ac.cr}
\affiliation    {School of Physics $^1$ and Space Research Center of the University of Costa Rica $^2$}
\date{\today}

\begin{abstract}
 An study of the equatorial circular motion of photons and massive particles around a rotating compact body
like a neutron star is presented. For this goal, we use an approximate Kerr-like metric with mass quadrupole
as perturbation. The effect of this deformation on the photon sphere, and the innermost stable circular orbit,
is determined via an effective potential. Furthermore, a stability anaysis is shown, where we observed the same
behavior of the effective potential for co-rotating and counter-rotating particles. A comparison with
the results for the innermost stable circular orbit obtained for the Hartle-Thorne spacetime is also presented.

\end{abstract}

\maketitle

\section{Introduction}

The study of the spacetime of a compact body with certain physical properties is built upon mathematical procedures deduced from the theory of General Relativity. These methods allow one to obtain information about the space-time surrounding, for example, a black hole or the Innermost Stable Circular Orbit (ISCO). The papers of Chandrasekhar and Bardeen \cite{Chandrasekhar, Bardeen} establish the appropriate formalisms that one can apply to a metric in order to describe, for the purpose of this research, the orbits around a compact body. 

Regarding the metric that will be employed in this research, Frutos \cite{Frutos, frutos2} implemented a Kerr spacetime as the seed to obtain a new one that incorporates the mass quadrupole with the aim of knowing its influence in the propagation of particles, such as photons. Moreover, one should go back to the Kerr case if the quadrupole parameter is null, which will be helpful further ahead to compare the results with a much simpler scenario. It is worthwhile to mention that applications of this spacetime, such as chaotic orbits, have already been studied in \cite{adrian, andre}.

It is important to mention that this contribution will try to part from various of the results from \cite{main}, where Pugliese et al. studied the equatorial circular motion for the Kerr metric. Furthermore, they found stability conditions in terms of the mass and spin parameters for a test particle orbiting a black hole and a naked singularity. Thus, one of the objectives of this research will be to determine, through numerical results, how the orbits change when the quadrupole parameter is added as a perturbation.

Additionally, the paper by Chaverri et al. \cite{ISCO} provides useful information about the Kerr-like metric in \cite{Frutos}, such as the effective potential and the radius for the ISCO, which will be analyzed in more detail in Section \ref{TF}. In fact, this research is an extension of what was accomplished in \cite{ISCO}, since the latter found the basic expressions for what will be analyzed in the present work.

In summary, the outline of this contribution is as follows: in the second section, the Kerr-like metric will be discussed.  As mentioned, the goal of the current paper is to analyze the influence of the quadrupole parameter on the massive and light orbits for the Kerr-like metric on the basis of numerical results generated by codes in Mathematica. Parallel to the latter, we will determine the radii of massive particles and photon orbits for this spacetime for different values of mass, rotation parameter, and quadrupole parameter, which is shown in section three. Additionally, the results for the ISCO will be briefly compared with those obtained with the Hartle-Thorne metric \cite{quevedo}. Moreover, in section four we present an analysis of the effect of the quadrupole parameter on the stability of circular orbits of test particles  considering their angular momentum through the study of the effective potential. All the numerical codes written in Mathematica employed for this task are available upon request. Lastly, some conclusions are discussed in section five.

\section{Kerr-like metric} \label{TF}
The Kerr-like metric is an approximate spacetime that was generated using the Kerr metric and the series expansion of the Erez-Rosen spacetime as seed metrics:

\begin{eqnarray} \label{met_Q}
    ds^2 =\frac{e^{-2\psi}}{\rho^2}(a^2 \sin^2\theta-\Delta)dt^2+\frac{-2Jr}{\rho^2} \sin^2\theta dtd\phi 
    + \rho^2\frac{e^{2\chi}}{\Delta}dr^2 
    + \rho^2e^{2\chi}d\theta^2 
    + \frac{e^{2\psi}}{\rho^2} [ (r^2+a^2)^2-a^2\Delta \sin^2\theta ] \sin^2\theta d\phi^2,
\end{eqnarray}

\noindent
where $\Delta = r^2-2Mr+a^2$, $\rho^2=r^2+a^2\cos^2\theta$; $J=Ma$ is the angular momentum; $M$ and $a$ correspond to the mass and the spin parameter, respectively. The quadrupole moment $q$ appears in the $\psi$ y $\chi$ functions as follows: 

\begin{eqnarray} \label{cuad1}
    \psi &=& \frac{q}{r^3}P_2+3\frac{Mq}{r^4}P_2 \\
     \label{cuad2}
    \chi &=& \frac{q}{r^3}P_2+\frac{1}{3}\frac{Mq}{r^4}(-1+5P_2+5P_2^2) \nonumber 
    +\frac{1}{9}\frac{q^2}{r^6}(2-6P_2-21P_2^2+25P_2^3)
\end{eqnarray} \noindent

\noindent and the metric components:

\begin{eqnarray}
    g_{tt}&=&\frac{e^{-2\psi}}{\rho^2}(a^2 \sin^2\theta-\Delta) \\
    g_{t\phi}&=&\frac{-2Jr}{\rho^2} \sin^2\theta \\
    g_{rr}&=&\rho^2\frac{e^{2\chi}}{\Delta} \\
    g_{\theta \theta}&=&\rho^2e^{2\chi} \\
    g_{\phi \phi}&=&\frac{e^{2\psi}}{\rho^2} [ (r^2+a^2)^2-a^2\Delta \sin^2\theta ] \sin^2\theta 
\end{eqnarray}

\noindent
where $P_2 = 1/2 (3 \cos^2{\theta} - 1)$. It is important to note that the metric returns to the Kerr case when $a=0$. Also, the Schwarzchild spacetime appears when $q=0$ and $a=0$ \cite{ISCO}.

As established in the Introduction, the metric used in this paper is an approximation,  but there are other methods that allow one to find the exact metric of a rotating compact body with mass quadrupole moment \cite{exact}. Nevertheless, one of the most recognized solutions to the Einstein equations in this particular case is the Hartle-Thorne metric, which is obtained by means of an approximation \cite{HTgeo, HTNS, neutS}.

\subsection{Hamiltonian}

The Lagrangian is defined as follows \cite{ISCO}: 

\begin{equation}
    L = \frac{\mu}{2} \frac{ds^2}{d\lambda}
\end{equation}
\noindent
where $\lambda$ is the affine parameter.

This expression allows to obtain the Hamiltonian by means of a Legendre transformation: 

\begin{equation} \label{hamiltonian}
    H = \frac{1}{2} (-E \dot{t} +g_{rr} \dot{r}^2 + L_z \dot{\phi}) = \epsilon
\end{equation}
\noindent

where $E$ and $L_z$ are the energy and angular momentum of the particles and are conserved quantities. A dot on a variable means derivative with respect to $\lambda$. Moreover, $\epsilon$ allows to specify the geodesic types.

\subsection{Effective potential and orbits}

From the Hamiltonian, it is possible to make a comparison with classical mechanics to find an effective potential, \cite{ISCO, hamveff, marek, sun}:

\begin{align}
\frac{V_{ef}}{\varepsilon} & = - 2 \bigg(1 - 2 M u + a^{2} u^{2} + q u^{3} 
- \frac{1}{2}{M q u^{4}} + \frac{5}{4} {q^{2} u^{6}} \bigg) 
+ \frac{L_{z}^{2}}{\varepsilon} \left(u^{2} + 2 q u^{5} + \frac{1}{2} {M q u^{6}} 
+\frac{11}{4} {q^{2} u^{8}} \right) 
- 2 M u^{3}(L_{z} - E a)^{2} \nonumber \\
& - \frac{E^{2}}{\varepsilon} \left(1 + a^{2} u^{2} - \frac{3}{4} {M q u^{4}} 
+ \frac{3}{4} {q^{2} u^{6}} \right) \label{peff} 
\end{align}

The effective potential gives information about the massive and light orbits. In the case of the latter, the numerical values can be obtained with the solutions of the denominator of the energy expression \cite{Bardeen, ISCO}:

\begin{align} 
\label{er}
\frac{E^2}{\varepsilon} & = \frac{2}{Z_{\mp}}
\bigg[(1 - 2 M u) \left(1 - 2 M u \pm 2 a u \sqrt{M u} \right) 
+ \left(a^2 M - \frac{1}{2} q \right) u^3 + \frac{25}{2} M q u^4 
+ \frac{29}{2} q^2 u^6 \bigg] 
\end{align}

\noindent

where

\begin{align} \label{z}
 Z_{\pm} & = \left(1 - 3 M u + \frac{33}{2} M q u^4 + 11 q^2 u^6 \right) 
\pm 2 a u \sqrt{M u} \textup{, } 
u = 1/r \textup{,} 
\nonumber \\
\end{align}

\noindent

The $\epsilon$ constant represents the Hamiltonian and it is set as 1 \cite{ISCO}.

The radii for the photon orbits ($r_{pht \pm}$) are the same as the roots of \ref{z}, since these values would cause a divergence in \ref{er}, which states that the orbit would have infinite energy per unit mass. This latter means that a photon could be "trapped" and be forced to orbit the compact body \cite{Bardeen, pht1, pht2, pht3, chaos}. 

As mentioned before, this Kerr-like metric could be applied to a rotating and deformed object, such as a neutron star. Although most of the latter do not have a photon sphere, an increment in the density (i.e. an "ultracompact" neutron star) would allow the light to orbit around said compact body \cite{NSpht}. 

There is one type of compact body usually taken into account in this type of analysis which is known as a Naked Singularity. The latter is characterized for the absence of an event horizon and the inability to "trap" light or make it orbit around it. This is still a theoretical object, since there is no experimental evidence for their existence \cite{noNS, NS1, NS2}. 

Apart from that, the angular momentum is an expression that will be helpful further ahead. It was also found in \cite{ISCO}:

\begin{align} 
\frac{L_z^2}{\varepsilon} & =\frac{2 u}{{\cal A}} \bigg[M - 3 M^2 u 
+ \left(2 M a^2 - \frac{3}{2} q \right) u^2 
+ \left(6 M^2 a^2 - \frac{5}{2} M q \right) u^3 + M \left(a^4 - 12 M^2 a^2\right) u^4 
+ \left(5 M^2 a^4 + 6 q^2\right) u^5 
\nonumber \\
& \pm 2 M a u \sqrt{M u} \big(a^4 u^4 - 2 M a^2 u^3 
+ 4 a^2 u^2 - 6 M u + 3\big) \bigg], \label{lr}
\end{align}

Additionally, it is important to mention the Inner most Stable Circular Orbit (ISCO), which is the inner edge of the accretion disk \cite{main, quevedo}. It is determined from the dynamics employing an effective potential and its derivatives to set specific conditions: 

\begin{equation} \label{cond_circ}
    \dot{r} = 0, \qquad V_{ef} = \frac{E}{\mu}, \qquad \frac{\partial V}{\partial r} = 0,
\end{equation}

For the analysis that we intend, it is also necessary to take into account the necessary conditions to make the motion equatorial: 

\begin{equation} \label{cond_circ}
    \theta = \frac{\pi}{2}, \qquad \frac{d\theta}{d\tau}=0,
\end{equation}

\noindent
where $\tau$ is the particles proper time.

The ISCO was found by Chaverri et al and it is associated with the following equation \cite{ISCO}:

\begin{align}
{\cal P} & = M r^5 - 9 M^2 r^4 
+ 3 \left(6 M^3 - M a^2 + \frac{1}{2} q \right) r^3 
- \left(7 M^2 a^2 - \frac{29}{2} M q \right) r^2 
- \frac{33}{2} q^2 \pm 6 M a r \sqrt{M r} \Delta = 0  \label{pisco}
\end{align}

\noindent

This equation can be scaled so the mass is not a free parameter anymore: $a\rightarrow a M$, $q\rightarrow q M^3$. With these modifications, we obtain the following expression, which allows to calculate numerical values of $r_{ISCO \pm}/M$:

\begin{align}
-3 a^2 \zeta ^3-7 a^2 \zeta ^2 \pm 12 a^2 \zeta ^{5/2} \pm6 a \zeta ^{3/2} \pm6 a \zeta ^{7/2} +\zeta ^5- 
9 \zeta ^4+18 \zeta ^3-\frac{33 q^2}{2}+\frac{3q}{2} \zeta ^3 +\frac{29 q}{2} \zeta ^2  = 0
\label{isco}
\end{align}
    
\noindent

where $\zeta=r/M$.

Before getting to equations \ref{er} and \ref{lr}, which expresses  the energy and the angular momentum in terms of the radius of the orbit, \cite{Frutos} allows to find an expression for $E$ also in terms of the $L_z$: 

\begin{equation}
    E^2 = \varepsilon (7 q^2 u^6-8 q u^4-q u^3-2 u+2) + x^2 u^3 (1 - 10 q u^3 - 33/4 q^2 u^5)
\end{equation}
\noindent
where $x = (a E - L_z)$. 

Once this equation is solved for $E$, it gives two solutions:

\begin{align}
\centering
\label{EL}
\frac{E_{\pm}}{\sqrt{\varepsilon}} &= \frac{1}{2 {\cal Q}^{-1}} \left\{ \frac{33}{2} a L_z q^2 u^8+20 a L_z q u^6-2 a L_z u^3 \pm 
\left[ \left(-\frac{33}{2} a L_z q^2 u^8-20 a L_z q u^6+2 a L_z u^3\right)^2 \right. \right. \\ \nonumber
& \left. \left. - 4 {\cal Q} \left(\frac{33}{4} L_z^2 q^2 u^8+10 L_z^2 q u^6-L_z^2 u^3-7 q^2 u^6+8 q u^4+q u^3+2 u-2\right) \right]^{1/2} \right\} , 
\end{align}

\noindent
where

$$ {\cal Q} = \frac{33}{4} a^2 q^2 u^8+10 a^2 q u^6-a^2 u^3+1 . $$

\begin{figure*}[h!]
\centering
\subfloat(a){\label{a}\includegraphics[width=8cm]{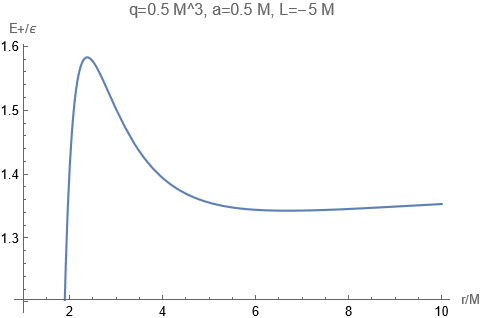}}
\subfloat(b){\label{b} \includegraphics[width=8cm]{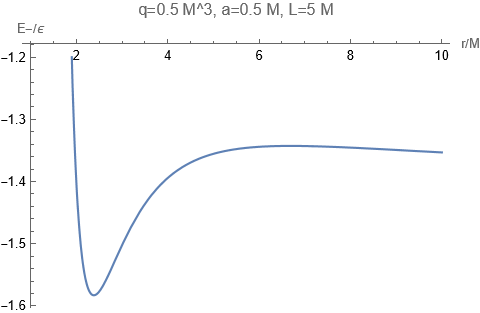}} \\
\vspace{1cm}
\subfloat(c){\label{c} \includegraphics[width=8cm]{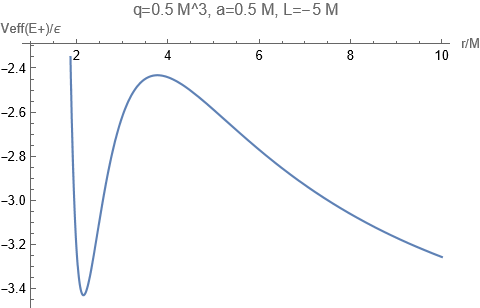}}
\subfloat(d){\label{d} \includegraphics[width=8cm]{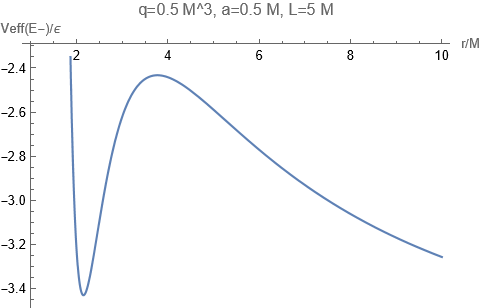}}
\caption{Energies for a positive/negative fixed angular momentum and its respective effective potential. Notice that the energies are opposite to each other, but the effective potential generated is exactly the same. (a) and (c) plot the case with $L_z<0$, while (b) and (d) represent $L_z>0$.}
\label{ex}
\end{figure*}

\begin{figure}
    \centering
    \includegraphics{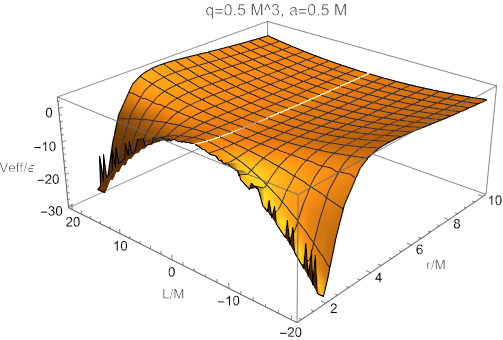}
    \caption{Surface produced by the effective potential for negative and positive values of the angular momentum.}
    \label{surex}
\end{figure}

The dependence of the energy with the angular momentum allows to fix a specific value of $L_{\pm}$ and study the orbits for co-rotating particles ($L_{-}$ and $E_{-}$) and counter-rotating particles ($-L_{+}$ and $E_{+}$) \cite{main}. Nevertheless, these combination of signs happen to give the same effective potential whenever the absolute value of the fixed angular momentum is the same, i.e., $\vert -L \vert$ = $\vert L \vert$. This can be visualized in the figure \ref{ex} and \ref{surex}, where it is also possible to see that both energies are exactly opposite to each other.

Lastly, algebraic software, such as Mathematica, will provide the possibility to perform complex calculations, plots and hence obtain numerical results that will be useful to understand the dependence between the ISCO and the fixed parameters. The Mathematica code used in the current research is available upon request.

\section{Numerical results for the photon orbits and ISCO radii}

In this section, the numerical results of the photon orbit and the ISCO when $a$ is constant and $q$ varies are shown in tables \ref{a1res}-\ref{a6res}, while tables \ref{q1res}-\ref{q6res} display the data obtained for $q$ fixed and different values of $a$. 

It should be clarified that equations \ref{z} and \ref{isco} might not have solutions for one of the two possible signs in the $\pm$ involved in the expressions. In that scenario, the tables mentioned before show a "-". Besides that, whenever there is more than one solution for the photon orbits and/or the ISCO, a coma separates the roots found for $Z$ and $P$. 

Moreover, since there is no expression for the Kerr-like metric event horizon and this paper does not intend to find one, the Kerr metric event horizon ($r_+=M+\sqrt{M^2-a^2}$) has been established as an approximate filter for the calculated radii \cite{Hioki}. In detail, these have to be higher than $r+$, since it would not make physical sense in the opposite case.

Let it be known that $r_{pht \pm}/M$ and $r_{ISCO \pm}/M$  with $q=0$ return to their respective values for the Schwarzschild metric. On the other hand, the Kerr metric results are obtained back in table \ref{a6res} when $q=0$. This proves that the Kerr-like spacetime is consistent with the existing theory, as expected. It also means that the numerical approximation done so far is correct.

\begin{table}[h!]
    \centering
    \caption{Values of $r_{pht+}/M$, $r_{pht-}/M$, $r_{ISCO+}/M$, and $r_{ISCO-}/M$ for different values of $q/M^3$ with $a/M=0.1$.}
    \label{a1res}
    \begin{tabular}{|c|c|c|c|c|}
        \hline
        $q/M^3$ & $r_{pht+}/M$ & $r_{pht-}/M$ & $r_{ISCO+}/M$ & $r_{ISCO-}/M$ \\
        \hline
           0.00 & 2.88219 & 3.11335 & 3.11335, 5.6693 & 2.88219, 6.32289 \\
        0.02 & 2.86789 & 3.10248 & 3.16314, 5.63646 & 2.91921, 6.30098 \\
        0.04 & 2.85308 & 3.09135 & 3.21358, 5.60248 & 2.95619, 6.27871 \\
        0.06 & 2.83774 & 3.07993 & 3.26483, 5.56727 & 2.99317, 6.25605 \\
        0.08 & 2.8218 & 3.06822 & 3.31703, 5.53068 & 3.0302, 6.23299 \\
        0.10 & 2.80521 & 3.05619 & 3.37039, 5.49254 & 3.06732, 6.2095 \\
        0.12 & 2.7879 & 3.04382 & 3.42512, 5.45266 & 3.10456, 6.18558 \\
        0.14 & 2.76979 & 3.03108 & 3.48149, 5.41079 & 3.14197, 6.16118 \\
        0.16 & 2.75079 & 3.01796 & 3.53982, 5.36662 & 3.17959, 6.13628 \\
        0.18 & 2.73078 & 3.00441 & 3.60054, 5.31976 & 3.21747, 6.11086 \\
        0.20 & 2.70961 & 2.99041 & 3.66418, 5.26966 & 3.25564, 6.08487 \\
        \hline
    \end{tabular}
\end{table}

\begin{table}[h!]
    \centering
    \caption{Values of $r_{pht+}/M$, $r_{pht-}/M$, $r_{ISCO+}/M$, and $r_{ISCO-}/M$ for different values of $q/M^3$ with $a/M=0.15$.}
    \label{a2res}
    \begin{tabular}{|c|c|c|c|c|}
        \hline
        $q/M^3$ & $r_{pht+}/M$ & $r_{pht-}/M$ & $r_{ISCO+}/M$ & $r_{ISCO-}/M$ \\
        \hline
        0.00 & 2.8214 & 3.16854 & 3.16854, 5.50062 & 2.8214, 6.48175 \\
        0.03 & 2.798 & 3.15313 & 3.25115, 5.4442 & 2.87377, 6.45143 \\
        0.06 & 2.77319 & 3.13718 & 3.33651, 5.38395 & 2.92596, 6.42044 \\
        0.09 & 2.74676 & 3.12066 & 3.42555, 5.31902 & 2.97807, 6.38874 \\
        0.12 & 2.71841 & 3.1035 & 3.51955, 5.24822 & 3.03021, 6.35629 \\
        0.15 & 2.68778 & 3.08566 & 3.62042, 5.16973 & 3.0825, 6.32303 \\
        0.18 & 2.65435 & 3.06705 & 3.73119, 5.08056 & 3.13503, 6.28891 \\
        0.21 & 2.61737 & 3.04759 & 3.85752, 4.97512 & 3.18792, 6.25384 \\
        0.24 & 2.57576 & 3.02718 & 4.01276, 4.84011 & 3.24129, 6.21777 \\
        0.27 & 2.52768 & 3.00571 & 4.25822, 4.61427 & 3.29526, 6.1806 \\
        0.30 & 2.46974 & 2.98302 & - & 3.34996, 6.14222 \\
        \hline
    \end{tabular}
\end{table}

\begin{table}[h!]
    \centering
    \caption{Values of $r_{pht+}/M$, $r_{pht-}/M$, $r_{ISCO+}/M$, and $r_{ISCO-}/M$ for different values of $q/M^3$ with $a/M=0.25$.}
    \label{a3res}
    \begin{tabular}{|c|c|c|c|c|}
        \hline
        $q/M^3$ & $r_{pht+}/M$ & $r_{pht-}/M$ & $r_{ISCO+}/M$ & $r_{ISCO-}/M$ \\
        \hline
       0.00 & 2.69545 & 3.27624 & 3.27624, 5.15554 & 2.69545, 6.79485 \\
        0.05 & 2.6481 & 3.25318 & 3.45534, 5.02333 & 2.77422, 6.75129 \\
        0.10 & 2.59411 & 3.22893 & 3.6655, 4.85676 & 2.85202, 6.70648 \\
        0.15 & 2.53058 & 3.20333 & 3.96191, 4.60114 & 2.92926, 6.66033 \\
        0.20 & 2.45182 & 3.1762 & - & 3.00628, 6.61271 \\
        0.25 & 2.34281 & 3.1473 & - & 3.08341, 6.56349 \\
        0.30 & - & 3.11631 & - & 3.16099, 6.51249 \\
        0.35 & - & 3.08282 & - & 3.23933, 6.45953 \\
        0.40 & - & 3.0463 & - & 3.31878, 6.40436 \\
        0.45 & - & 3.00594 & - & 3.39972, 6.34672 \\
        0.50 & - & 2.96057 & - & 3.48259, 6.28624 \\
        \hline
    \end{tabular}
\end{table}

\begin{table}[h!]
    \centering
    \caption{Values of $r_{pht+}/M$, $r_{pht-}/M$, $r_{ISCO+}/M$, and $r_{ISCO-}/M$ for different values of $q/M^3$ with $a/M=0.5$.}
    \label{a4res}
    \begin{tabular}{|c|c|c|c|c|}
        \hline
        $q/M^3$ & $r_{pht+}/M$ & $r_{pht-}/M$ & $r_{ISCO+}/M$ & $r_{ISCO-}/M$ \\
        \hline
             0.00 & 2.3473 & 3.53209 & 3.53209, 4.233 & 2.3473, 7.55458 \\
        0.08 & 2.20069 & 3.50343 & - & 2.45293, 7.50383 \\
        0.16 & - & 3.47302 & - & 2.55418, 7.45174 \\
        0.24 & - & 3.44057 & - & 2.65232, 7.39824 \\
        0.32 & - & 3.40575 & - & 2.74827, 7.34319 \\
        0.40 & - & 3.36809 & - & 2.8428, 7.28647 \\
        0.48 & - & 3.32696 & - & 2.93651, 7.22792 \\
        0.56 & - & 3.28147 & - & 3.02997, 7.16736 \\
        0.64 & - & 3.2303 & - & 3.12367, 7.10459 \\
        0.72 & - & 3.17127, 1.97216 & - & 3.2181, 7.03935 \\
        0.80 & - & 3.10046, 2.10603 & - & 3.31376, 6.97134  \\
        \hline
    \end{tabular}
\end{table}

\begin{table}[h!]
    \centering
    \caption{Values of $r_{pht+}/M$, $r_{pht-}/M$, $r_{ISCO+}/M$, and $r_{ISCO-}/M$ for different values of $q/M^3$ with $a/M=0.7$.}
    \label{a5res}
    \begin{tabular}{|c|c|c|c|c|}
        \hline
        $q/M^3$ & $r_{pht+}/M$ & $r_{pht-}/M$ & $r_{ISCO+}/M$ & $r_{ISCO-}/M$ \\
        \hline
        0.00 & 2.01333 & 3.72535 & 3.39313, 3.72535 & 2.01333, 8.14297 \\
        0.09 & - & 3.69849 & - & 2.12761, 8.09665 \\
        0.18 & - & 3.67017 & - & 2.23337, 8.04941 \\
        0.27 & - & 3.64017 & - & 2.33313, 8.00119 \\
        0.36 & - & 3.60825 & - & 2.42854, 7.95192 \\
        0.45 & - & 3.57409 & - & 2.52073, 7.90155 \\
        0.54 & - & 3.53727 & - & 2.61055, 7.84999 \\
        0.63 & - & 3.49726, 1.71532 & - & 2.69869, 7.79716 \\
        0.72 & - & 3.45328, 1.83434 & - & 2.78566, 7.74297 \\
        0.81 & - & 3.40422, 1.95271 & - & 2.87194, 7.68732 \\
        0.90 & - & 3.34834, 2.0732 & - & 2.95792, 7.63009 \\
        \hline
    \end{tabular}
\end{table}

\begin{table}[h!]
    \centering
    \caption{Values of $r_{pht+}/M$, $r_{pht-}/M$, $r_{ISCO+}/M$, and $r_{ISCO-}/M$ for different values of $q/M^3$ with $a/M=1$.}
    \label{a6res}
    \begin{tabular}{|c|c|c|c|c|}
        \hline
        $q/M^3$ & $r_{pht+}/M$ & $r_{pht-}/M$ & $r_{ISCO+}/M$ & $r_{ISCO-}/M$ \\
        \hline
        0.0 & - & 4.0 & 1.0, 4.0 & 1.0, 9.0 \\
        0.1 & - & 3.97659 & 2.66322, 3.42996 & 1.38284, 8.96036 \\
        0.2 & - & 3.95212 & - & 1.54836, 8.92017 \\
        0.3 & - & 3.92649, 1.15365 & - & 1.67818, 8.87941 \\
        0.4 & - & 3.89954, 1.30341 & - & 1.78997, 8.83807 \\
        0.5 & - & 3.87113, 1.43659 & - & 1.89061, 8.7961 \\
        0.6 & - & 3.84105, 1.55899 & - & 1.98366, 8.75349 \\
        0.7 & - & 3.80904, 1.67409 & - & 2.07124, 8.71021 \\
        0.8 & - & 3.7748, 1.78428 & - & 2.15475, 8.66622 \\
        0.9 & - & 3.73791, 1.89139 & - & 2.23518, 8.62149 \\
        1.0 & - & 3.69783, 1.99695 & 1.0148 & 1.01488, 2.31325, 8.57599 \\
        \hline
    \end{tabular}
\end{table}

\begin{table}[h!]
    \centering
    \caption{Values of $r_{pht+}/M$, $r_{ISCO-}/M$, $r_{ISCO+}/M$, and $r_{pht-}/M$ for different values of $a/M$ with $q/M^3 = 0.1$.}
    \label{q1res}
    \resizebox{8cm}{!}{
    \begin{tabular}{|c|c|c|c|c|}
    \hline
    $a/M$ & $r_{pht+}/M$ & $r_{pht-}/M$ & $r_{ISCO+}/M$ & $r_{ISCO-}/M$ \\
    \hline
       0.00 & 2.93418 & 2.93418 & 3.21399, 5.86141 & 3.21399, 5.86141 \\
    0.02 & 2.90899 & 2.95909 & 3.2442, 5.78961 & 3.18415, 5.93241 \\
    0.04 & 2.88352 & 2.98373 & 3.27484, 5.71692 & 3.15462, 6.00266 \\
    0.06 & 2.85774 & 3.00812 & 3.30602, 5.64325 & 3.12533, 6.07223 \\
    0.08 & 2.83164 & 3.03227 & 3.33782, 5.56851 & 3.09625, 6.14116 \\
    0.10 & 2.80521 & 3.05619 & 3.37039, 5.49254 & 3.06732, 6.2095 \\
    0.12 & 2.77842 & 3.07988 & 3.40387, 5.4152 & 3.03851, 6.27729 \\
    0.14 & 2.75127 & 3.10335 & 3.43847, 5.33628 & 3.00978, 6.34456 \\
    0.16 & 2.72371 & 3.12662 & 3.47442, 5.25552 & 2.98111, 6.41134 \\
    0.18 & 2.69574 & 3.14969 & 3.51206, 5.1726 & 2.95246, 6.47766 \\
    0.20 & 2.66732 & 3.17256 & 3.55184, 5.08705 & 2.92381, 6.54355  \\
    \hline
    \end{tabular}
    }
\end{table}

\begin{table}[h!]
    \centering
    \caption{Values of $r_{pht+}/M$, $r_{pht-}/M$, $r_{ISCO+}/M$, and $r_{pht-}/M$ for different values of $a/M$ with $q/M^3 = 0.15$.}
    \label{q2res}
    \begin{tabular}{|c|c|c|c|c|}
    \hline
    $a/M$ & $r_{pht+}/M$ & $r_{pht-}/M$ & $r_{ISCO+}/M$ & $r_{ISCO-}/M$ \\
    \hline
    0.00 & 2.89699 & 2.89699 & 3.32487, 5.78489 & 3.32487, 5.78489 \\
    0.03 & 2.85708 & 2.9361 & 3.37738, 5.67039 & 3.27416, 5.89663 \\
    0.06 & 2.8163 & 2.97447 & 3.43224, 5.55259 & 3.22485, 6.00603 \\
    0.09 & 2.77456 & 3.01215 & 3.4902, 5.43072 & 3.17663, 6.11341 \\
    0.12 & 2.73176 & 3.0492 & 3.55236, 5.30367 & 3.12925, 6.21901 \\
    0.15 & 2.68778 & 3.08566 & 3.62042, 5.16973 & 3.0825, 6.32303 \\
    0.18 & 2.64248 & 3.12155 & 3.69726, 5.02598 & 3.03621, 6.42565 \\
    0.21 & 2.59569 & 3.15693 & 3.78854, 4.86676 & 2.99025, 6.52699 \\
    0.24 & 2.54718 & 3.19181 & 3.90853, 4.67777 & 2.94449, 6.62718 \\
    0.27 & 2.49666 & 3.22623 & 4.14279, 4.37341 & 2.89881, 6.7263 \\
    0.30 & 2.44375 & 3.2602 & - & 2.85312, 6.82445 \\
    \hline
    \end{tabular}
\end{table}

\begin{table}[h!]
    \centering
    \caption{Values of $r_{pht+}/M$, $r_{pht-}/M$, $r_{ISCO+}/M$, and $r_{ISCO-}/M$ for different values of $a/M$ with $q/M^3 = 0.25$.}
    \label{q3res}
    \begin{tabular}{|c|c|c|c|c|}
    \hline
    $a/M$ & $r_{pht+}/M$ & $r_{pht-}/M$ & $r_{ISCO+}/M$ & $r_{ISCO-}/M$ \\
    \hline
       0.00 & 2.81021 & 2.81021 & 3.56187, 5.61122 & 3.56187, 5.61122 \\
    0.05 & 2.73291 & 2.88336 & 3.68701, 5.38402 & 3.45248, 5.82023 \\
    0.10 & 2.65031 & 2.95316 & 3.84179, 5.12471 & 3.35271, 6.01721 \\
    0.15 & 2.56057 & 3.02015 & 4.07694, 4.78254 & 3.25921, 6.20551 \\
    0.20 & 2.46042 & 3.08476 & - & 3.1699, 6.38715 \\
    0.25 & 2.34281 & 3.1473 & - & 3.08341, 6.56349 \\
    0.30 & 2.18458 & 3.20802 & - & 2.99873, 6.73548 \\
    0.35 & - & 3.26714 & - & 2.91508, 6.9038 \\
    0.40 & - & 3.32481 & - & 2.83184, 7.06898 \\
    0.45 & - & 3.38118 & - & 2.74845, 7.23142 \\
    0.50 & - & 3.43636 & - & 2.66441, 7.39144 \\
    \hline
    \end{tabular}
\end{table}

\begin{table}[h!]
    \centering
    \caption{Values of $r_{pht+}/M$, $r_{pht-}/M$, $r_{ISCO+}/M$, and $r_{ISCO-}/M$ for different values of $a/M$ with $q/M^3 = 0.5$.}
    \label{q4res}
    \begin{tabular}{|c|c|c|c|c|}
    \hline
    $a/M$ & $r_{pht+}/M$ & $r_{pht-}/M$ & $r_{ISCO+}/M$ & $r_{ISCO-}/M$ \\
    \hline
    0.00 & - & - & - & - \\
    0.08 & 2.61521, 2.0056 & - & - & 3.99199, 5.46916 \\
    0.16 & 2.79907 & - & - & 3.71694, 5.89209 \\
    0.24 & 2.94399 & - & - & 3.50651, 6.24492 \\
    0.32 & 3.07007 & - & - & 3.32459, 6.56353  \\
    0.40 & 3.18443 & - & - & 3.15771, 6.86107 \\
    0.48 & 3.29054 & - & - & 2.99888, 7.14404 \\
    0.56 & 3.39044 & - & - & 2.84368, 7.41617 \\
    0.64 & 3.48543 & - & - & 2.68875, 7.67985 \\
    0.72 & 3.5764 & - & - &  2.531, 7.93669 \\
    0.80 & 3.66399 & - & - & 2.36705, 8.18786 \\
    \hline
    \end{tabular}
\end{table}

\begin{table}[h!]
    \centering
    \caption{Values of $r_{pht+}/M$, $r_{pht-}/M$, $r_{ISCO+}/M$, and $r_{ISCO-}/M$ for different values of $a/M$ with $q/M^3 = 0.7$.}
    \label{q5res}
    \begin{tabular}{|c|c|c|c|c|}
    \hline
    $a/M$ & $r_{pht+}/M$ & $r_{pht-}/M$ & $r_{ISCO+}/M$ & $r_{ISCO-}/M$ \\
    \hline
     0.00 & - & - & - & - \\
    0.09 & - & - & - & - \\
    0.18 & - & - & - & 4.12482, 5.60428 \\
    0.27 & - & 2.71234, 2.24479 & - & 3.78036, 6.10441 \\
    0.36 & - & 2.94153, 2.08255 & - & 3.52575, 6.50789 \\
    0.45 & - & 3.10683, 1.98357 & - & 3.30733, 6.86781 \\
    0.54 & - & 3.24704, 1.90908 & - & 3.10671, 7.20179 \\
    0.63 & - & 3.37268, 1.84855 & - & 2.9144, 7.51816 \\
    0.72 & - & 3.48848, 1.79729 & - & 2.72396, 7.82163 \\
    0.81 & - & 3.59705, 1.75272 & - & 2.52972, 8.11513 \\
    0.90 & - & 3.69998, 1.71325 & - & 2.32492, 8.40065 \\
    \hline
    \end{tabular}
\end{table}

\begin{table}[h!]
    \centering
    \caption{Values of $r_{pht+}/M$, $r_{pht-}/M$, $r_{ISCO+}/M$, and $r_{ISCO-}/M$ for different values of $a/M$ with $q/M^3 = 1$.}
    \label{q6res}
    \begin{tabular}{|c|c|c|c|c|}
    \hline
    $a/M$ & $r_{pht+}/M$ & $r_{pht-}/M$ & $r_{ISCO+}/M$ & $r_{ISCO-}/M$ \\
    \hline
    0 & - & - & - & - \\
    0.1 & - & - & - & - \\
    0.2 & - & - & - & - \\
    0.3 & - & - & - & 4.31446, 5.72417 \\
    0.4 & - & - & - & 3.87274, 6.32523 \\
    0.5 & - & - & - & 3.56194, 6.78664 \\
    0.6 & - & 3.07871, 2.34011 & - & 3.29754, 7.19168 \\
    0.7 & - & 3.27451, 2.2142 & - & 3.05356, 7.56447 \\
    0.8 & - & 3.43254, 2.12545 & - & 2.81617, 7.91587 \\
    0.9 & - & 3.57131, 2.05536 & - & 2.5742, 8.25183 \\
    1 & - & 3.69783, 1.99695 & 1.0148 & 1.01488, 2.31325, 8.57599 \\
    \hline
    \end{tabular}
    
\end{table}

It is possible to see that, whenever $a$ is fixed, the increase of the quadrupole parameter causes a contraction in the photon orbits and the ISCO, except for the cases where a "degeneracy" of certain radii appears. Nevertheless, there is still a tendency for each component of these pairs of radii: one of them is always smaller than its companion, but grows along with the quadrupole parameter; the other is, in consequence, bigger, but decreases as $q$ increases. The exact opposite happens when $q$ is fixed and $a$ magnifies.  

On the other hand, whenever $q$ is fixed, the increase of the spin parameter causes an expansion in the (-) cases of the photon orbits and the ISCO. The same degeneracy mentioned before manifests for the case $q/M^3=0.7$, and there is a specific tendency for each of the radius in both pairs: the smaller one decreases as $q$ increases and the bigger one grows along with $q$.

Although it is important to recognize that there exists a radius with an opposite tendency for some combination of parameters, the results indicate that $q$ reduces the distance at which an accretion disk can be formed for a compact body. Furthermore, the faster the object rotates, the larger these orbits are. 

Lastly, there is a decrease on the amount of photon orbits and solutions of the ISCO the faster the compact body rotates. There does not seem to be a clear pattern or limit for the parameters that cause this effect. Nevertheless, it is more common for high values of $a$ and big differences between $a$ and $q$. 

The results obtained for the ISCO can be compared to those found in \cite{quevedo} for the Hartle-Thorne metric. The latter plots a unique solution for the ISCO radius as a function of the spin and quadrupole parameters. 

It is clear that the increase of $j$ in \cite{quevedo} causes the ISCO to be smaller, while the opposite happens the larger $q$ is. These tendencies are followed by one of the degenerate roots of the ISCO, whether it is the (+) or the (-) case. Nevertheless, its companion radius behaves in exactly the opposite way. Therefore, for each sign there is one root that obeys the pattern given by the Hartle-Thorne metric and another one that does exactly the opposite.  

\section{Stability}

For this section, the effective potential is plotted for different combinations of $a$ and $q$. Given that the surfaces are exactly the same in both positive and negative halves of the angular momentum axis and to improve visualization, it is only shown the co-rotative part of the effective potential. The behavior shown in the latter should be the same as in the counter-rotative region (for an opposite value of $L_z$).  

Moreover, 2D plots that show the effective potential against the radius are selected from the surface so it is possible to find radial intervals (delimited by the ISCO's and photon orbits) and angular momentum limits for the stable and unstable orbits associated to a specific combination of parameters. Let it be clear that the minima and maxima points necessary for the analysis just described are found with the Mathematica functions FindMinimum and FindMaximum. The seed is selected after taking a look at the respective 2D plots. The AccuracyGoal and PrecissionGoal are fixed in Automatic because the main interest is to know whether there is a minimum/maximum or not and to obtain an approximate radial location.

As said before, the angular momentum will be fixed in order to study the orbits with the effective potential. The values of $L_z$ are chosen in order to picture the intervals in the tables mentioned before.

Once again, the Kerr event horizon is settled as an approximated limit to the results, given that we are not interested in considering data from a region that cannot contain orbits for particles or light.

\subsection{General}

\begin{figure}[h!]
    \centering
    \includegraphics{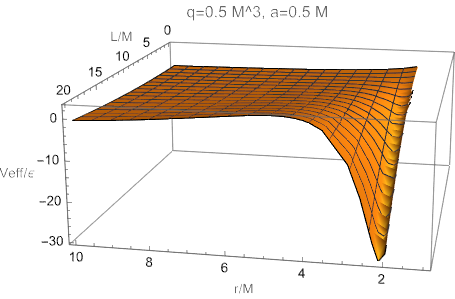}
    \caption{Effective potential with $a=0.5M$ and $q=0.5M^3$.}
    \label{sura=q}
\end{figure}

\begin{figure}
    \centering
    \includegraphics{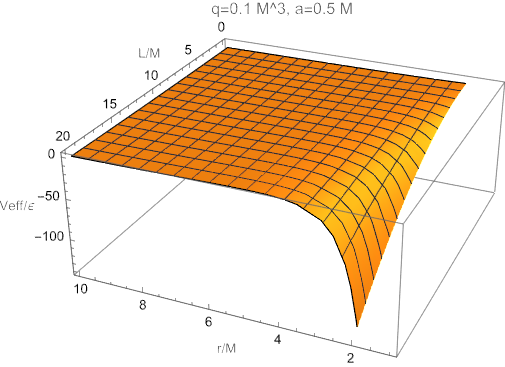}
    \caption{Effective potential with $a=0.5M$ and $q=0.1M^3$.}
    \label{surabq}
\end{figure}

\begin{figure}[h!]
    \centering
    \includegraphics{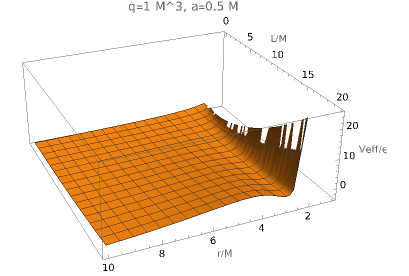}
    \caption{Effective potential with $a=0.5M$ and $q = M^3$.}
    \label{suralq}
\end{figure}

First, three basic cases are analysed: $a$ equal to $q$ (figure \ref{sura=q}), $a$ higher than $q$ (figure \ref{surabq}), and $a$ lower than $q$ (figure \ref{suralq}). 

\begin{figure*}[h!]
\centering
\subfloat(a){\label{a}\includegraphics[width=8cm]{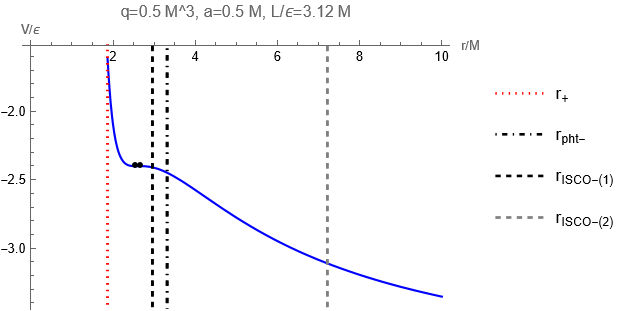}}
\subfloat(b){\label{b} \includegraphics[width=8cm]{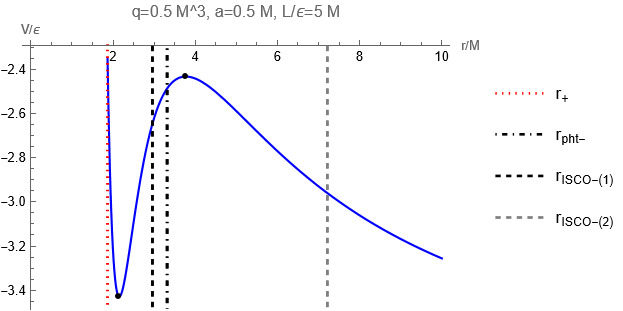}} \\
\vspace{1cm}
\subfloat(c){\label{c} \includegraphics[width=8cm]{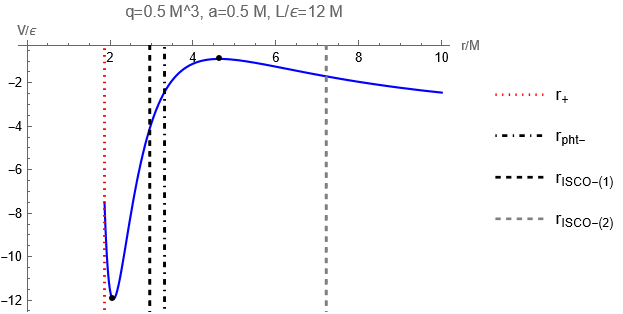}}
\caption{Effective potential for different values of $L_z$ with $a=0.5M$ and $q=0.5M^3$. Maxima (unstable orbits): (a) at (2.66274, -2.40098), with $E_-/\varepsilon = -1.29984$, (b) at (3.77188, -2.43275), with $E_-/\varepsilon = -1.41057$, (c) at (4.63761, -0.901079) with $E_-/\varepsilon = -1.77053$. Minima (stable orbits): (a) at (2.55292, -2.40124) with $E_-/\varepsilon = -1.29589$,(b) at (2.15138, -3.42892) with $E_-/\varepsilon = -1.53804$, (c) at (2.07731, -11.957) with $E_-/\varepsilon = -2.9431$.}
\label{a=q}
\end{figure*}

For the first scenario, both stable and unstable orbits can be found. As it is possible to see in the plots from figure \ref{a=q}, the increase in the angular momentum causes the minima and maxima to be further from the compact body. 

In the $a$ lower than $q$ case, only unstable orbits are produced, and these tent to be more external for higher angular momentum values. 

Finally, whenever $q$ is larger than $a$, the results are similar to those from the first case, but the change in $q$ (now larger) causes the stable/unstable orbits to begin at higher angular momentum values. Also, the increase in $L_z$ has the same influence on the maxima, but it has an opposite effect on the minima: the higher the angular momentum is, the closer the stable orbits are from the compact body. 

\begin{table}[h!]
\begin{tabular}{l|cc|}
\cline{2-3}
                               & \multicolumn{2}{c|}{$q = 0.5 M^3$, $a = 0.5 M$}                       \\ \cline{2-3} 
                               & \multicolumn{1}{c|}{Angular momentum interval} & Radial region        \\ \hline
\multicolumn{1}{|c|}{Stable}   & \multicolumn{1}{c|}{$L_z > 3.19 M$}            & $r < r_{ISCO - (2)}$ \\ \hline
\multicolumn{1}{|c|}{Unstable} & \multicolumn{1}{c|}{$L_z > 3.19 M$}            & $r < r_{ISCO - (2)}$ \\ \hline
\end{tabular}
\caption{Possible angular momentum values and radial regions for stable and unstable orbits in the effective potential with 
$q = 0.5 M^3$ and $a = 0.5 M$.}
\label{tab:a=q}
\end{table}

\begin{figure*}[h!]
\centering
\subfloat(a){\label{a}\includegraphics[width=8cm]{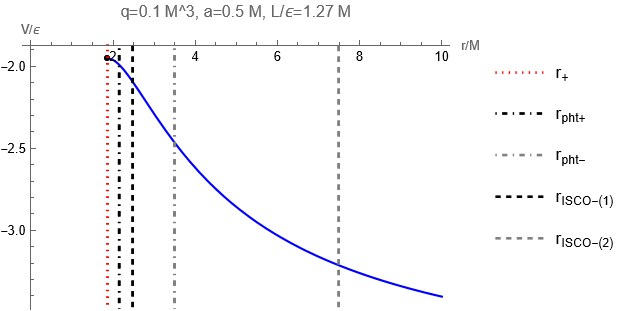}}
\subfloat(b){\label{b} \includegraphics[width=8cm]{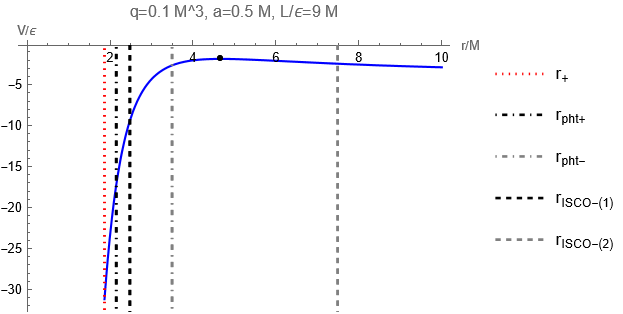}} \\
\caption{Effective potential for different values of $L_z$ with $a=0.5M$ and $q=0.1M^3$. Maxima (unstable orbits): (a) at (1.87126, -1.9528), with $E_-/\varepsilon = -1.13446$, (b) at (4.67241, -1.85804) with $E_-/\varepsilon = -1.58106$.}
\label{abq}
\end{figure*}

\begin{table}[h!]
\begin{tabular}{l|cc|}
\cline{2-3}
                               & \multicolumn{2}{c|}{$q = 0.1 M^3$, $a = 0.5 M$}                       \\ \cline{2-3} 
                               & \multicolumn{1}{c|}{Angular momentum interval} & Radial region        \\ \hline
\multicolumn{1}{|c|}{Stable}   & \multicolumn{1}{c|}{-}                         & -                    \\ \hline
\multicolumn{1}{|c|}{Unstable} & \multicolumn{1}{c|}{$L_z > 1.28 M$}            & $r < r_{ISCO - (2)}$ \\ \hline
\end{tabular}
\caption{Possible angular momentum values and radial regions for stable and unstable orbits in the effective potential with 
$q = 0.1 M^3$ and $a = 0.5 M$.}
\label{tab:abq}
\end{table}

\begin{figure*}[h!]
\centering
\subfloat(a){\label{a}\includegraphics[width=8cm]{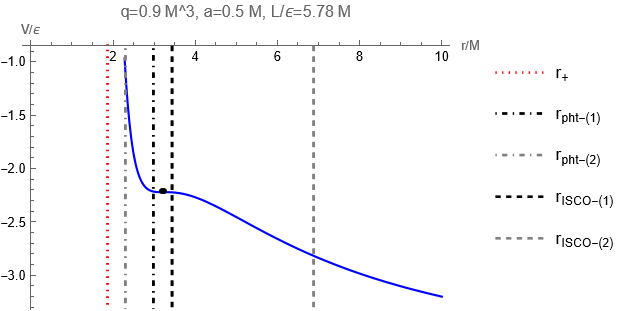}}
\subfloat(b){\label{b} \includegraphics[width=8cm]{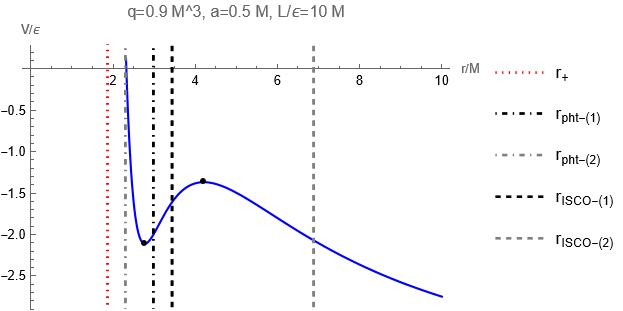}} \\
\vspace{1cm}
\subfloat(c){\label{c} \includegraphics[width=8cm]{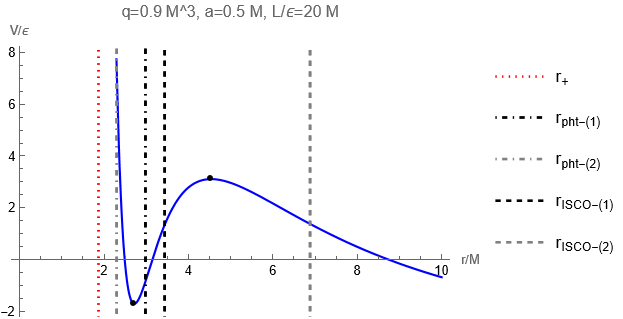}}
\caption{Effective potential for different values of $L_z$ with $a=0.5M$ and $q=M^3$. Maxima (unstable orbits): (a) at (3.26388, -2.22094), with $E_-/\varepsilon = -1.60876$ (b) at (4.20912, -1.36905) with $E_-/\varepsilon = -1.75685$, (c) at (4.53867, 3.10496) with $E_-/\varepsilon = -2.5176$. Minima (stable orbits): (a) at (3.20423, -2.22097) with $E_-/\varepsilon = -1.62671$, (b) at (2.77345, -2.11063) with $E_-/\varepsilon = -2.64278$, (c) at (2.71622, -1.71677) with $E_-/\varepsilon = -5.02585$.}
\label{alq}
\end{figure*}

\begin{table}[h!]
\begin{tabular}{l|cc|}
\cline{2-3}
                               & \multicolumn{2}{c|}{$q = 0.9 M^3$, $a = 0.5 M$}                                     \\ \cline{2-3} 
                               & \multicolumn{1}{c|}{Angular momentum interval} & Radial region                      \\ \hline
\multicolumn{1}{|c|}{Stable}   & \multicolumn{1}{c|}{$L_z > 5.98 M$}            & $r_{pht - (2)}<r < r_{ISCO - (1)}$ \\ \hline
\multicolumn{1}{|c|}{Unstable} & \multicolumn{1}{c|}{$L_z > 5.98 M$}            & $r_{pht - (1)}<r < r_{ISCO - (2)}$ \\ \hline
\end{tabular}
\caption{Possible angular momentum values and radial regions for stable and unstable orbits in the effective potential with 
$q = 0.9 M^3$ and $a = 0.5 M$.}
\label{tab:alq}
\end{table}

\subsection{Extreme case}

In this subsection, the value of $a$ is set as equal to $M$. The main difference with the results seen until now is that, despite $q$ always being lower than $a$, it is possible to find stable orbits, which sit closer to the compact body with the increase in angular momentum. In addition to this, the tables \ref{tab:x1}, \ref{tab:x2}, and \ref{tab:x3} make evident that the increase of the quadrupole parameter causes the minimum $L_z$ value for minima/maxima to be higher. 

\begin{figure}[h!]
    \centering
    \includegraphics{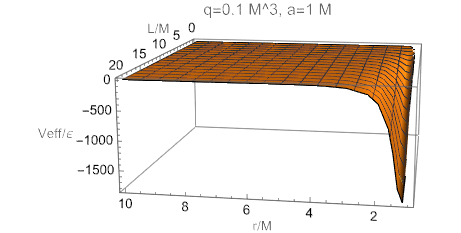}
    \caption{Effective potential with $a=M$ and $q = 0.1M^3$.}
    \label{surx1}
\end{figure}

\begin{figure*}[h!]
\centering
\subfloat(a){\label{a}\includegraphics[width=8cm]{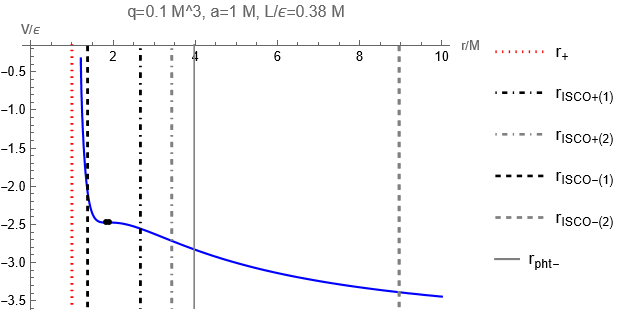}}
\subfloat(b){\label{b} \includegraphics[width=8cm]{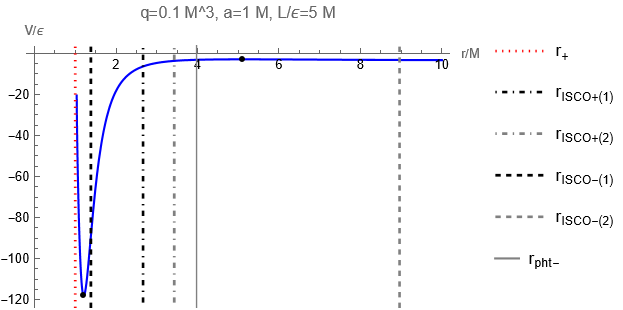}} \\
\caption{Effective potential for different values of $L_z$ with $a=M$ and $q=0.1M^3$. Maxima (unstable orbits): (a) at (1.92869, -2.47565), with $E_-/\varepsilon = -1.06946$, (b) at (5.11728, -2.92928) with $E_-/\varepsilon = -1.38163$. Minima (stable orbits): (a) at (1.83596, -2.47593) with $E_-/\varepsilon = -1.0495$, (b) at (1.20812, -117.884) with $E_-/\varepsilon = -4.54246$.}
\label{x1}
\end{figure*}

\begin{table}[h!]
\begin{tabular}{l|cc|}
\cline{2-3}
                               & \multicolumn{2}{c|}{$q = 0.1 M^3$, $a = M$}                                        \\ \cline{2-3} 
                               & \multicolumn{1}{c|}{Angular momentum interval} & Radial region                       \\ \hline
\multicolumn{1}{|c|}{Stable}   & \multicolumn{1}{c|}{$L_z > 0.39 M$}            & $r < r_{ISCO + (1)}$                \\ \hline
\multicolumn{1}{|c|}{Unstable} & \multicolumn{1}{c|}{$L_z > 0.39 M$}            & $r_{ISCO - (1)}<r < r_{ISCO - (2)}$ \\ \hline
\end{tabular}
\caption{Possible angular momentum values and radial regions for stable and unstable orbits in the effective potential with 
$q = 0.1 M^3$ and $a = M$.}
\label{tab:x1}
\end{table}

\begin{figure}[h!]
    \centering
    \includegraphics{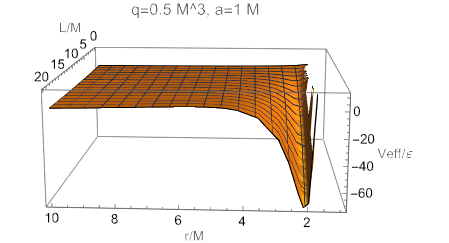}
    \caption{Effective potential with $a=M$ and $q = 0.5M^3$.}
    \label{surx2}
\end{figure}

\begin{figure*}[h!]
\centering
\subfloat(a){\label{a}\includegraphics[width=8cm]{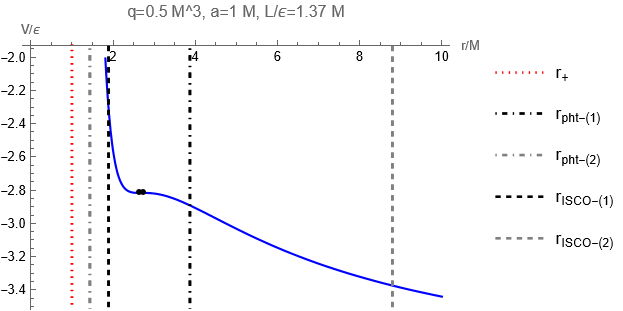}}
\subfloat(b){\label{b} \includegraphics[width=8cm]{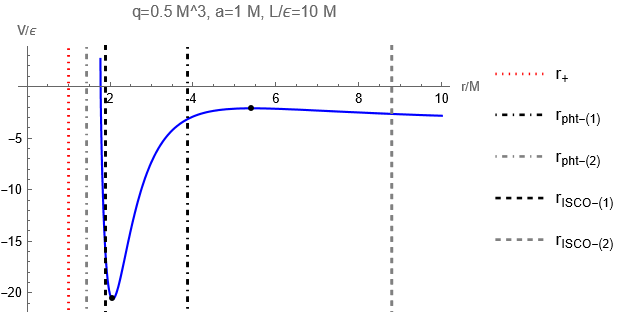}} \\
\caption{Effective potential for different values of $L_z$ with $a=M$ and $q=0.5M^3$. Maxima (unstable orbits): (a) at (2.73843, -2.81681), with $E_-/\varepsilon = -1.18972$, (b) at (5.42236, -2.10592) with $E_-/\varepsilon = -1.56047$. Minima (stable orbits): (a) at (2.64327, -2.81689) with $E_-/\varepsilon = -1.17918$, (b) at (2.06342, -20.6084) with $E_-/\varepsilon = -2.78731$.}
\label{x2}
\end{figure*}

\begin{table}[]
\begin{tabular}{l|cc|}
\cline{2-3}
                               & \multicolumn{2}{c|}{$q = 0.5 M^3$, $a = M$}                                        \\ \cline{2-3} 
                               & \multicolumn{1}{c|}{Angular momentum interval} & Radial region                       \\ \hline
\multicolumn{1}{|c|}{Stable}   & \multicolumn{1}{c|}{$L_z > 1.37 M$}            & $ r_{ISCO - (1)}<r$                 \\ \hline
\multicolumn{1}{|c|}{Unstable} & \multicolumn{1}{c|}{$L_z > 1.37 M$}            & $r_{ISCO - (1)}<r < r_{ISCO - (2)}$ \\ \hline
\end{tabular}
\caption{Possible angular momentum values and radial regions for stable and unstable orbits in the effective potential with 
$q = 0.5 M^3$ and $a = 1 M$.}
\label{tab:x2}
\end{table}

\begin{figure}[h!]
    \centering
    \includegraphics{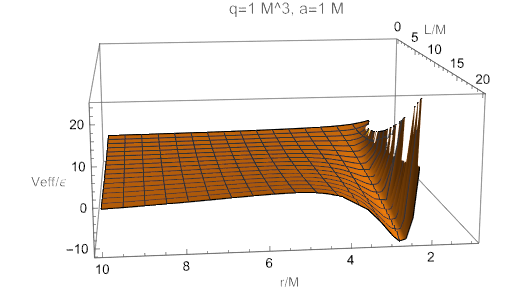}
    \caption{Effective potential with $a=M$ and $q = M^3$.}
    \label{surx3}
\end{figure}

\begin{figure*}[h!]
\centering
\subfloat(a){\label{a}\includegraphics[width=8cm]{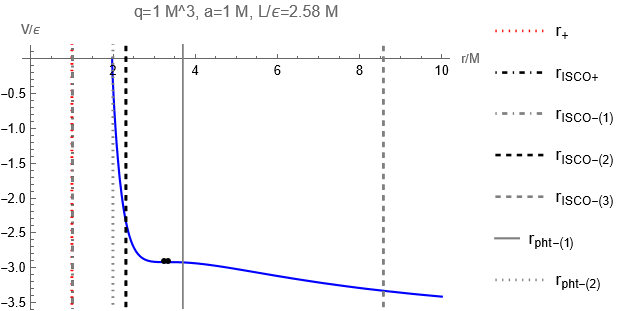}}
\subfloat(b){\label{b} \includegraphics[width=8cm]{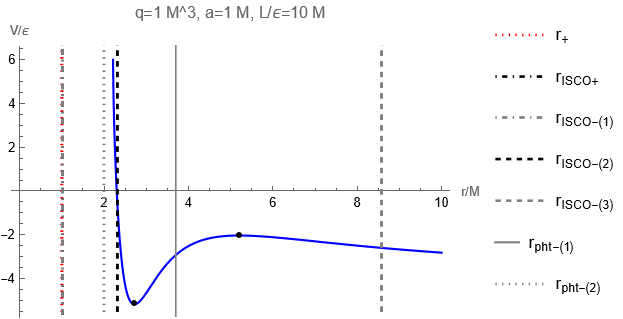}} \\
\caption{Effective potential for different values of $L_z$ with $a=M$ and $q=M^3$. Maxima (unstable orbits): (a) at (3.36211, -2.92171), with $E_-/\varepsilon = -1.26501$, (b) at (5.20639, -2.03959) with $E_-/\varepsilon = -1.57422$. Minima (stable orbits): (a) at (3.2585, -2.92178) with $E_-/\varepsilon = -1.25917$, (b) at (2.72116, -5.1865) with $E_-/\varepsilon = -2.08216$.}
\label{x3}
\end{figure*}

\begin{table}[h!]
\begin{tabular}{l|cc|}
\cline{2-3}
                               & \multicolumn{2}{c|}{$q = M^3$, $a = M$}                                              \\ \cline{2-3} 
                               & \multicolumn{1}{c|}{Angular momentum interval} & Radial region                       \\ \hline
\multicolumn{1}{|c|}{Stable}   & \multicolumn{1}{c|}{$L_z > 2.58 M$}            & $ r_{ISCO - (2)}<r<r_{pht - (1)}$   \\ \hline
\multicolumn{1}{|c|}{Unstable} & \multicolumn{1}{c|}{$L_z > 2.58M$}             & $r_{ISCO - (2)}<r < r_{ISCO - (3)}$ \\ \hline
\end{tabular}
\caption{Possible angular momentum values and radial regions for stable and unstable orbits in the effective potential with 
$q = M^3$ and $a = 1 M$.}
\label{tab:x3}
\end{table}

\subsection{Schwarzchild and Kerr cases:}

\begin{figure}[h!]
    \centering
    \includegraphics{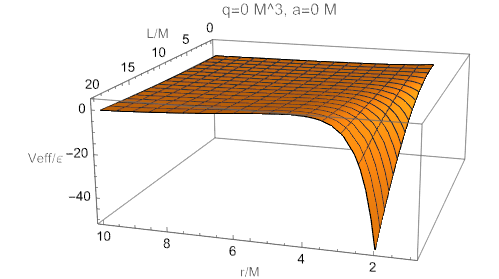}
    \caption{Effective potential with $a/M=0$ and $q/M^3=0$.}
    \label{surkrr1}
\end{figure}

\begin{figure}[h!]
    \centering
    \includegraphics{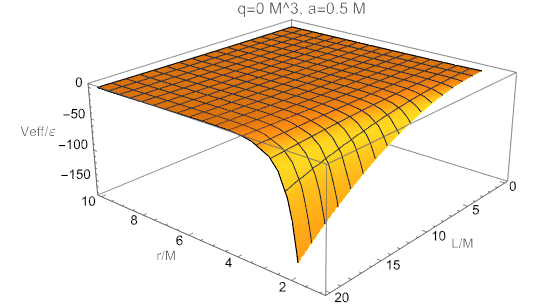}
    \caption{Effective potential with $a=0.5M$ and $q/M^3=0$.}
    \label{surkrr2}
\end{figure}

\begin{figure}
    \centering
    \includegraphics{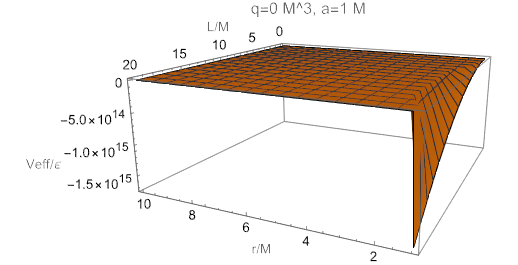}
    \caption{Effective potential with $a=M$ and $q/M^3=0$.}
    \label{surkrr3}
\end{figure}

Lastly, the Schwarzchild ($a/M=0, q/M^3=0$) and Kerr ($q/M^3$) cases are plotted in figures \ref{surkrr1}, \ref{surkrr2} and \ref{surkrr3}. The most important detail out of these three scenarios is that none of them present stable orbits. Also, the increase of the angular momentum causes the maxima to be further from the compact body. 

Contrary to what was determined for the relation between the quadrupole parameter and the lowest angular momentum value to produce a minimum/maximum in the other cases, the higher the spin parameter, the smaller the angular parameter magnitude requirement is. An example of this is that, for the surface with $q=M^3$ and $a=M$ it was found that every $L_z$ value generates an unstable orbit in a valid radial interval, which is no delimited by any ISCO or photon orbit (that is why there is not a table for the analysis of the surface).

\begin{figure*}[h!]
\centering
\subfloat(a){\label{a}\includegraphics[width=8cm]{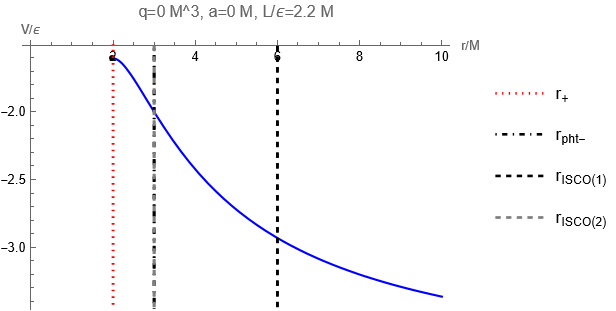}}
\subfloat(b){\label{b} \includegraphics[width=8cm]{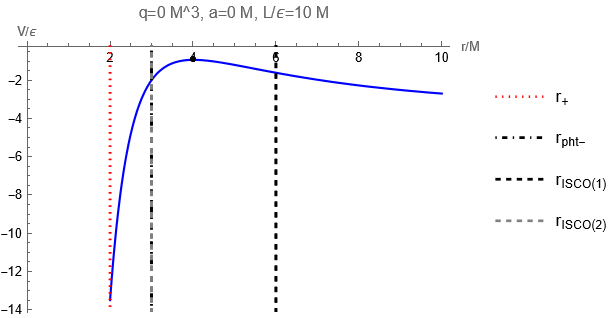}} \\
\caption{Effective potential for different values of $L_z$ with $a/M=0$ and $q/M^3=0$. Maxima (unstable orbits): (a) at (2.00593, -1.60495) with $E_-/\varepsilon = -1.26594$, (b) at (4.01612, -0.937375) with $E_-/\varepsilon = -1.74521$.}
\label{krr1}
\end{figure*}

\begin{table}[]
\begin{tabular}{l|cc|}
\cline{2-3}
                               & \multicolumn{2}{c|}{$q / M^3  = 0$, $a / M = 0$}                    \\ \cline{2-3} 
                               & \multicolumn{1}{c|}{Angular momentum interval} & Radial region      \\ \hline
\multicolumn{1}{|c|}{Stable}   & \multicolumn{1}{c|}{-}                         & -                  \\ \hline
\multicolumn{1}{|c|}{Unstable} & \multicolumn{1}{c|}{$L_z > 2.20M$}             & $r < r_{ISCO (1)}$ \\ \hline
\end{tabular}
\caption{Possible angular momentum values and radial regions for stable and unstable orbits in the effective potential with 
$q / M^3 = 0$ and $a / M = 0$.}
\label{tab:krr1}
\end{table}

\begin{figure*}[h!]
\centering
\subfloat(a){\label{a}\includegraphics[width=8cm]{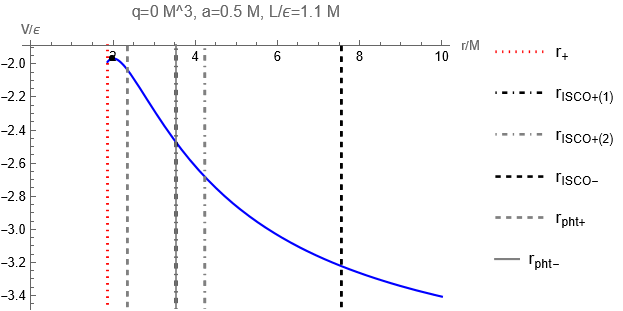}}
\subfloat(b){\label{b} \includegraphics[width=8cm]{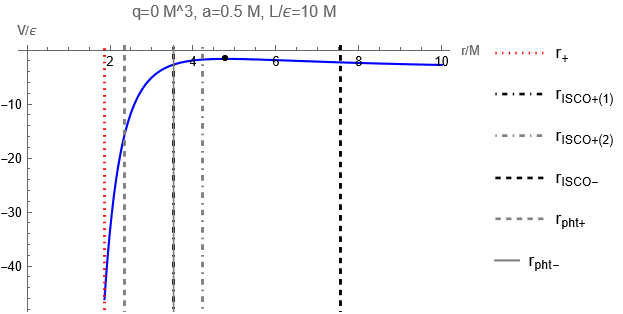}} \\
\caption{Effective potential for different values of $L_z$ with $a=0.5M$ and $q/M^3=0$. Maxima (unstable orbits): (a) at (2.00511, -1.96762) with $E_-/\varepsilon$ = -1.16334, (b) at (4.78924, -1.63142) with $E_-/\varepsilon$ = -1.62692.}
\label{krr2}
\end{figure*}

\begin{table}[h!]
\begin{tabular}{l|cc|}
\cline{2-3}
                               & \multicolumn{2}{c|}{$q / M^3 = 0$, $a = 0.5 M$}                   \\ \cline{2-3} 
                               & \multicolumn{1}{c|}{Angular momentum interval} & Radial region    \\ \hline
\multicolumn{1}{|c|}{Stable}   & \multicolumn{1}{c|}{-}                         & -                \\ \hline
\multicolumn{1}{|c|}{Unstable} & \multicolumn{1}{c|}{$L_z > 1.10M$}             & $r < r_{ISCO -}$ \\ \hline
\end{tabular}
\caption{Possible angular momentum values and radial regions for stable and unstable orbits in the effective potential with 
$q / M^3 = 0$ and $a = 0.5 M$.}
\label{tab:krr2}
\end{table}

\begin{figure*}[h!]
\centering
\subfloat(a){\label{a}\includegraphics[width=8cm]{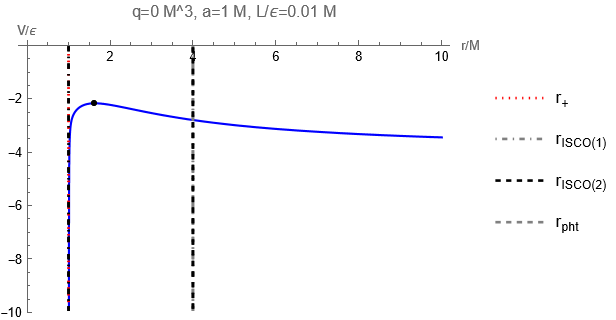}}
\subfloat(b){\label{b} \includegraphics[width=8cm]{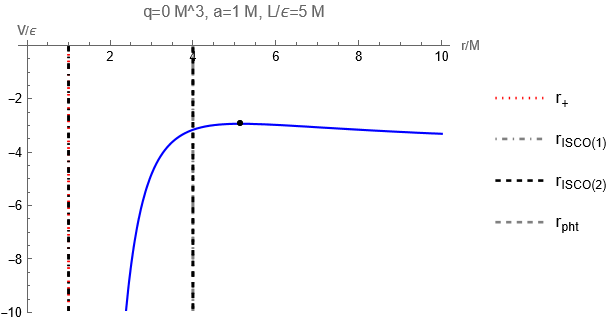}} \\
\caption{Effective potential for different values of $L_z$ with $a=M$ and $q/M^3=0$. Maxima (unstable orbits): (a) at (1.63464, -2.16675), with $E_-/\varepsilon = -1.00651$, (b) at (5.15135, -2.93433) with $E_-/\varepsilon = -1.38192$.}
\label{krr3}
\end{figure*}

\clearpage

\section{Conclusions}

The numerical results for the radii show that the increase of $q$ causes a shrinkage of the photon sphere, while $a$ tends to expand it. Therefore, the spin and quadrupole parameters have opposite effects on the light orbits of the Kerr-like metric. 

Furthermore, the ISCO can be degenerate in both + and - cases, which can even get to be triple, as it can be observed in tables \ref{a6res} and \ref{q6res}. Whenever there are two values per sign, the - case causes the smaller root to grow along with $q$ and the bigger one to decrease, but the opposite happens as $a$ gets larger. For the + case, the tendency is the same despite the parameter that is being increased: the bigger root grows and the smaller one shrinks.

On the other hand, the analysis of the effective potential shows that the increase in $q$ makes the angular momentum requirement higher for stable/unstable orbits, while the increase in $a$ does exactly the opposite. Moreover, the maxima locate further away from the compact body the larger the angular momentum ($L_z$) is, but there is not an exact tendency for the minima, as some become more external and others more  internal. 

It is also worthwhile to mention that the extreme case ($a=M$) allows us to obtain stable orbits even when the quadrupole parameter is lower than the spin, which did not happen in any other scenario seen. 

Moreover, when we compare the ISCO given by the Hartle-Thorne metric with our results, it is possible to appreciate that one component of the degenerate radius for each sign follows the behavior seen in \cite{quevedo}, while the other changes in exactly the opposite way whenever one of the parameters (spin and mass quadrupole) changes.  

In terms of future works, we would like to expand this numerical analysis to more complex types of spacetime and compare the results obtained in this contribution with these. For example, we could take the Kerr-like spacetime, add a magnetic dipole to it, and analyse its stability along with the photon orbits and the ISCO \cite{mag1, mag2, mag3}. Besides that, it would also be possible to study the shadow of a rotating compact body with mass quadrupole moment or even expand the analysis of the chaotic behavior in its orbits \cite{adrian}.

\end{document}